\documentclass[10pt,aps,prd,amsfonts,nofootinbib,supercriptaddress]{revtex4}
\usepackage{amssymb,amsmath}
\usepackage{graphicx}
\usepackage{epsfig}
\usepackage{bbm}

\newcounter{fig}

\newcommand{\beq}{\begin{equation}}
\newcommand{\eeq}{\end{equation}}
\newcommand{\bea}{\begin{eqnarray}}
\newcommand{\eea}{\end{eqnarray}}

\begin{document}



\title{Planck scale operators, inflation and fine tuning}

\author{Anja Marunovi\'c$\footnote{anja.marun@gmail.com}$}
\author{Tomislav Prokopec$^{a}\footnote{t.prokopec@uu.nl}$}
\affiliation{
 $^a$Institute for Theoretical Physics, Spinoza
Institute and EMMEPh, Utrecht University, Princetnlaan 5, 3584 CC Utrecht, The
Netherlands}

\begin{abstract} \noindent

 Ultraviolet completion of the standard model plus gravity at and beyond the Planck scale
is a daunting problem to which no generally accepted solution exists.
Principal obstacles include (a) lack of data at the Planck scale (b) nonrenormalizability of gravity
and (c) unitarity problem. Here we make a simple observation that, if one treats all
Planck scale operators of equal canonical
dimension democratically, one can tame some of the undesirable features of these models.
With a reasonable amount of fine tuning  one can satisfy slow roll conditions required in viable inflationary models.
That remains true even when the number of such operators becomes very large.
\end{abstract}


\maketitle

\section{Gravity as an effective theory}
\label{Gravity as effective theory}

 Arguably the simplest way of dealing with gravity below the Planck scale is to treat it as an effective theory,
according to which threshold effects from the unknown Planck scale theory
at scales sufficiently below the Planck scale can be subsumed to local operators~\cite{Donoghue:1994dn,Weinberg:2009bg}. 
Planck scale operators can be generated both by the threshold effects from Planck scale physics, and 
by the quantum effects of gravitational and matter fields from sub-Planckian scales.
A simple application of gravity as an effective theory results 
in quantum corrections to gravitational potentials,
which are unobservably small on Minkowski 
background~\cite{BjerrumBohr:2002kt,Marunovic:2012pr,Marunovic:2011zw} but can be much larger on 
curved backgrounds (on de Sitter background these corrections can be observably large~\cite{Park:2015kua}).
In other words, significantly below the Planck scale 
the effective action of gravity plus matter admits a gradient expansion. 
\footnote{The well known working example which Donoghue~\cite{Donoghue:1994dn} 
quotes is the non-linear sigma model of mesons, 
which is valid significantly below the scale $\Lambda_{\rm QCD}$, the scale of confinement at which QCD strongly 
couples and below which every perturbative calculation ceases to be valid.}
Assuming covariance,
this then limits the theory to a set of operators which is finite when truncated at any finite canonical dimension.
Since we are primarily interested in understanding inflation, here we focus on scalar-tensor theories,
for which the fundamental fields are the metric $g_{\mu\nu}$ and a scalar field $\phi$, which is 
in inflationary literature known as the inflaton.~\footnote{Some inflationary models comprise more scalar fields. 
Our considerations 
are easily generalized to encompass these more general cases. Similarly, our considerations can be 
generalized to include operators that include other (fermionic and vector) matter fields.} 

 The fundamental assertion of the effective theory approach to quantum gravity is that operators are classified according 
to their canonical (mass) dimension, and generally operators of lower dimension are more important.
 Here they are, ordered by their canonical dimension,
\begin{itemize}
\item[$(A)$] Dimension $d=0$: 1 (also known as the cosmological constant).
\item[$(B)$] $d=2$: $R$ (the Hilbert-Einstein term), $\phi^2$ (the scalar mass term).
\item[$(C)$] $d=4$: $R^2$, $R_{\mu\nu}R^{\mu\nu}$, $\phi^2R$ (the non-miniminal coupling of scalar to gravity), 
$G\!\! B$, $\phi^4$ and $(\partial_\mu\phi)(\partial^\mu\phi)$ (the scalar kinetic term).
\item[$(D)$] $d=6$: $R^3$, $RR_{\mu\nu}R^{\mu\nu},\dots$,\ $\phi^4R$, $\phi^2R^2$, $\phi^6$,  
$\phi^2(\partial_\mu\phi)(\partial^\mu\phi)$, $R(\partial_\mu\phi)(\partial^\mu\phi),$ 
$R^{\mu\nu}(\partial_\mu\phi)(\partial_\nu\phi),$
{\it etc.},
\end{itemize}
where $R$ and $R_{\mu\nu}$ denote the Ricci scalar and tensor, respectively, 
$G\!\! B=R^2-4R_{\mu\nu}R^{\mu\nu}+R_{\mu\nu\rho\sigma}R^{\mu\nu\rho\sigma}$ 
stands for the Gauss-Bonnet term, which is in four space-time dimensions a total derivative, and hence it 
can be discarded as it does not affect the equations of motion.
Note that the number of operators increases rapidly with increasing canonical dimension. 
Above we list all operators up to $d=4$ and give a sample of them in $d=6$. 
Operators of odd dimension also exist, for example $\phi R$ is a dimension 3 operator. Here for simplicity we assume that 
they are not present. One way of excluding them is symmetry. When $\phi$ is a real scalar,
the corresponding symmetry is $O(1)\cong \mathbbm{Z}_2$, which requires the action to be invariant under $\phi\rightarrow -\phi$.
In the case of $N$ real scalars, the corresponding symmetry is 
$O(N)$.~\footnote{When an internal symmetry is imposed on the classical level, it typically also survives at the quantum level, 
which is what we assume is the case in this work. Notable exceptions are spontaneous symmetry breaking 
by a Higgs-like condensate and `breaking' of gauge symmetry {\it via} quantum effects such as 
in the scalar electrodynamics on de Sitter space~\cite{Prokopec:2002jn,Prokopec:2002uw}.
}

 According to the effective theory of gravity, operators of lower dimensionality are more important.
If that does not happen, it creates a problem that begs for a theoretical explanation. For example, the fact that 
the $d=0$ operator is not important throughout the evolution of the universe (except possibly 
in the last seven billions of years) is known as the cosmological constant 
problem, and it is deemed as the most important unsolved hierarchy problem in modern physics~\cite{Weinberg:1988cp}.

Analogously, during inflation the coefficient of the $\phi^2$ term is anomalously small. A possible explanation is 
conformal symmetry that forbids such a term. The smallness of it can be then explained by a small violation of 
conformal symmetry generated by quantum effects. The mass scale of the Hilbert-Einstein term $R$ is the Planck 
scale $M_{\rm P}^2$. It sets the strength of coupling of gravity to matter (since $M_{\rm P}$ is large, its coupling 
to matter $\propto 1/M_{\rm P}^2$ is weak), and we have no idea why its value is what it is (for a recent discussion on that 
question see {\it e.g.} Refs.~\cite{Lucat:2016eze} and~\cite{Henz:2016aoh}).

 The first time that an operator that violates unitarity appears is at $d=4$. 
This is the $R_{\mu\nu}R^{\mu\nu}$ operator~\cite{Stelle:1976gc,Stelle:1977ry},
and when included into the graviton propagator, it generates a massive ghost field 
(this is  a field whose creation contributes negatively to the energy of the system), 
thereby destabilizing the theory~\cite{Sbisa:2014pzo,Boulware:1973my}.
The operator $R_{\mu\nu}R^{\mu\nu}$ destabilizes gravity even classically {\it via} the Ostrogradsky instability~\cite{Woodard:2015zca},
and therefore ought to be excluded. In what follows, we assume that operators that violate unitarity 
are not present in the effective theory, {\it i.e.} that they are excluded by some 
unknown mechanism operative at the Planck scale.
However, even if such a mechanism is present at the Planck scale, 
because gravity is a nonrenormalizable theory~\cite{'tHooft:1974bx,Goroff:1985th},  
these type of operators will be 
generated as a result of quantum fluctuations of matter fields on sub-Planckian scales. 
The only way of taming these operators is to assume 
that the resulting effective theory is applicable up to an energy scale sufficiently below the Planck scale, such that these 
operators do not wreck havoc in the Universe (recall that the ghost mass is Planckian, which suppresses its creation below 
the Planck scale).

 An obvious application of the present work is to the question of 
fine tuning in inflationary models~\cite{Miao:2015oba}.
 One often makes use of an approximate shift symmetry --
which is approximately respected by the tree-level cosmological perturbations --  
and then uses it as an argument to forbid all higher dimensional operators that violate that same shift symmetry.
It is however unclear why would such a symmetry be respected by the Planck scale physics~\cite{Mazumdar:2010sa}.
 One notable attempt where such a justification is argued
 is axion monodromy inflation~\cite{Silverstein:2008sg,McAllister:2008hb},
in which there are Goldstone bosons associated with a global symmetry that is softly broken
by the exponentially suppressed instanton corrections.
This is so because perturbative diagrams generate only derivative couplings to the Goldstone bosons. 
Some supersymmetric models of inflation~\cite{Dvali:1994ms,Binetruy:1996xj}
were initially conceived to avoid fine tuning problems of usual inflationary models.
However,  these type of models tend to suffer from severe tuning problems
when Planck scale higher dimensional operators are added~\cite{Dine:1995uk,Dine:1995kz,Enqvist:2003gh}. 
The role of quantum corrections and Planck scale operators in Higgs inflation has resulted in an extensive discussion
in the literature~\cite{Barbon:2009ya,Bezrukov:2009db,Burgess:2010zq,George:2013iia,George:2015nza}
(see also~\cite{Prokopec:2014iya}), 
leading to the conclusion that -- unless one forbids the Planck scale operators that violate shift symmetry --
the slow roll conditions of Higgs inflation will be 
quite generically destroyed~\cite{Bezrukov:2010jz}.
At the moment it is unclear however, why such operators should be forbidden.
In passing we note that the reasoning advocated in this work can be fruitfully applied to studying
the threshold and quantum corrections in Higgs inflation. 
The role of the one-loop inflaton quantum corrections in Higgs inflaton deserves a closer inspection 
and we intend to address it elsewhere.

 The remainder of the paper is divided into two parts. In section~\ref{Dimension four operators}
we discuss dimension four operators,
which is followed by discussion of dimension six operators in section~\ref{Dimension six operators},
in which we also remark on the role of operators of dimension $d >6$.
Some concluding remarks are given in section~\ref{Conclusions and discussion},

\section{Dimension four operators}
\label{Dimension four operators}

  In this section we consider the effective action for gravity and matter that includes all operators up to canonical
 dimension 4. For simplicity, for matter we take a real scalar field $\phi$ and for gravity we take all operators
 that are consistent with unitarity and we assume that the effective action is consistent with certain symmetries.
 In our case these are the general covariance of the graviton field and the $O(1)\cong {\mathbbm{Z}_2}$ symmetry 
of the real scalar.
The effective action in four space-time dimensions~\footnote{
In this paper we are not interested in studying quantum effects of matter and gravitational fields on the effective theory,
and therefore we present our results in $D=4$ space-time dimensions. A generalization of our considerations to general $D$ dimensions 
-- which is a good starting point for studying quantum effects by the method of dimensional regularization -- is straightforward.} 
is then,
\begin{equation}
 S[g_{\mu\nu},\phi]= \int d^4x\sqrt{-g}\left\{
    \frac{M_{\rm P}^2}{2}(R-\Lambda) +\frac{\alpha}{2}R^2-\frac{\xi}{2}R\phi^2
     -\frac12g^{\mu\nu}(\partial_\mu\phi)(\partial_\nu\phi)-\frac{m^2}{2}\phi^2-\frac{\lambda}{4!}\phi^4
 \right\}
\,,
\label{action dim 4}
\end{equation}
where $M_{\rm P}=1/[8\pi G_N]^{1/2}\simeq 2.4\times 10^{18}~{\rm GeV}$ 
is the reduced Planck mass, $G_N$ is the Newton constant, $\Lambda$ is the cosmological constant,
$m$ is the scalar field mass and $\alpha$, $\xi$ and $\lambda$ are dimensionless couplings of operators of 
canonical dimension $d=4$.
Here we work in units in which the speed of light $c=1$ and the reduced Planck constant $\hbar=h/(2\pi)=1$.
The action~(\ref{action dim 4}) is equivalent to 
the action,~\footnote{One can easily show the on-shell equivalence of the two actions as follows.
Varying Eq.~(\ref{action dim 4:multiplier}) with respect to $\omega$ and $\Phi$ gives,
$\omega(R-\Phi)=0$, $F^\prime-\omega^2=0$,
$F(\Phi)=\frac{\alpha}{2}\Phi^2+\frac{M_{\rm P}^2}{2}(\Phi-\Lambda)-\frac{\xi}{2}\Phi\phi^2$. 
The non-singular solutions of these equations are,
$\Phi=R$ (for $\omega\neq 0$) and $\omega^2=F^\prime(\Phi)$. Inserting these solutions into
the action~(\ref{action dim 4:multiplier}) gives Eq.~(\ref{action dim 4}), completing the proof of 
the on-shell equivalence of the actions~(\ref{action dim 4}) and~(\ref{action dim 4:multiplier}), which suffices for 
our purpose.
 }
\begin{equation}
 S = \int d^4x\sqrt{-g}\left\{
    \frac{M_{\rm P}^2}{2}(\Phi-\Lambda) +\frac{\alpha}{2}\Phi^2+\omega^2(R-\Phi)-\frac{\xi}{2}\Phi\phi^2
     -\frac12g^{\mu\nu}(\partial_\mu\phi)(\partial_\nu\phi)-\frac{m^2}{2}\phi^2-\frac{\lambda}{4!}\phi^4
 \right\}
\,,
\label{action dim 4:multiplier}
\end{equation}
where we have introduced a new scalar field $\Phi$ and a Lagrange multiplier (constraint) field $\omega=\omega(x)$.

Now upon varying the action~(\ref{action dim 4:multiplier}) with respect to $\Phi$ and solving the resulting equation, one
obtains,
\begin{equation}
 \Phi=\frac{1}{\alpha}\left[\omega^2+\frac{\xi}{2}\phi^2-\frac{M_{\rm P}^2}{2}\right]
\,.
\label{solution for Phi}
\end{equation}
Inserting this into~(\ref{action dim 4:multiplier}) results in an equivalent action,
\begin{equation}
 S = \int d^4x\sqrt{-g}\left\{
    \omega^2 R -\frac{M_{\rm P}^2}{2}\Lambda
     -\frac12g^{\mu\nu}(\partial_\mu\phi)(\partial_\nu\phi)-\frac{m^2}{2}\phi^2-\frac{\lambda}{4!}\phi^4
     -\frac{1}{2\alpha}\left[\omega^2+\frac{\xi}{2}\phi^2-\frac{M_{\rm P}^2}{2}\right]^2
 \right\}
\,.
\label{action dim 4:multiplier:2}
\end{equation}
In this action the Langrange multiplier $\omega$ appears as a non-minimally coupled scalar. It is hence useful to
transform to the Einstein frame by making use of a suitable conformal transformation, $g_{\mu\nu}=\Omega^2(x)g_{\mu\nu}^E$,
where $\Omega$ is a local (possibly field dependent) conformal function. By making use of the standard
conformal transformation for the Ricci scalar and determinant of the metric,
we get (our metric signature is $(-,+,+,+)$) (see {\it e.g.} Ref.~\cite{Dabrowski:2008kx}),
\begin{eqnarray}
 S &=& \int d^4x\sqrt{-g_E}\Bigg\{
    \Omega^{2}\omega^2 \left[R_E-6g_E^{\mu\nu}\frac{\nabla_\mu^E\nabla_\nu^E\Omega}{\Omega}
 -\Omega^{2}\frac12g_E^{\mu\nu}(\partial_\mu\phi)(\partial_\nu\phi)
           \right]
\nonumber\\
     && \hskip 2.2cm
-\Omega^4\frac{M_{\rm P}^2}{2}\Lambda
    -\Omega^4\frac{m^2}{2}\phi^2-\Omega^4\frac{\lambda}{4!}\phi^4
     -\frac{\Omega^4}{2\alpha}\left[\omega^2+\frac{\xi}{2}\phi^2-\frac{M_{\rm P}^2}{2}\right]^2
 \Bigg\}
\,.
\label{action dim 4:multiplier:3}
\end{eqnarray}
The correct choice for $\Omega$ that gets rid of the non-minimal coupling is,
\begin{equation}
\Omega^2=\frac{M_{\rm P}^2}{2}\frac{1}{\omega^{2}}
\,.
\label{Jordan to Einstein frame}
\end{equation}
Upon partially integrating the second term inside the square brackets in the first line of Eq.~(\ref{action dim 4:multiplier:3})
and dropping the resulting (presumably irrelevant) boundary term and upon using~(\ref{Jordan to Einstein frame}), 
Eq.~(\ref{action dim 4:multiplier:3}) simplifies to (for convenience we keep $\Omega$ in the action),
\begin{eqnarray}
 S &=& \int d^4x\sqrt{-g_E}\Bigg\{
     \frac{M_{\rm P}^2}{2}R_E
                   -3M_{\rm P}^2g_E^{\mu\nu}\frac{(\partial_\mu\Omega)(\partial_\nu\Omega)}{\Omega^2}
                   -\Omega^{2}\frac12g_E^{\mu\nu}(\partial_\mu\phi)(\partial_\nu\phi)
\label{action dim 4:multiplier:4}\\
     &&  \hskip 2.2cm
     -\Omega^4\frac{M_{\rm P}^2}{2}\Lambda
    -\Omega^4\frac{m^2}{2}\phi^2
      -\Omega^4\frac{\lambda}{4!}\phi^4
     -\frac{M_{\rm P}^4}{8\alpha}\left[1+\frac{\xi}{2}\frac{\phi^2}{M_{\rm P}^2}\Omega^{2}-\Omega^{2}\right]^2
 \Bigg\}
\,.
\nonumber
\end{eqnarray}
We have thus got rid of the higher dimensional gravitational operator in the original action~(\ref{action dim 4}), 
but the prize to pay is the emergence of
a second dynamical scalar known as the scalaron~\cite{Starobinsky:1980te}.
Note that both scalars in~(\ref{action dim 4:multiplier:4})
 have non-canonical kinetic terms. The following simple rescaling brings the scalar kinetic terms
into their canonical form,
\begin{equation}
 \psi_E = \sqrt{6} M_{\rm P}\times\ln(\Omega)
\,,\qquad \phi_E = \Omega\phi        \;\Longrightarrow\; d\phi_E = \Omega d\phi
\,,
\label{Jordan to Einstein frame:2}
\end{equation}
where we choose the integration constants such that, when $\Omega=1$ and $\phi=0$ then $\psi_E=0$ and $\phi_E=0$
and we choose the positive root in the first field transformation in~(\ref{Jordan to Einstein frame:2})
(choosing the negative root would lead to a completely equivalent action, which is related
to the action obtained below by the transformation $\psi_E\rightarrow -\psi_E$).
The implication in~(\ref{Jordan to Einstein frame:2}) holds since $\Omega$ and $\phi$ can be 
considered as two independent fields (see Eq.~(\ref{tilde psiE}) below). With this 
action~(\ref{action dim 4:multiplier:4}) becomes the following Einstein frame action,
\begin{eqnarray}
 S_E[\psi_E,\phi_E] &=& \int d^4x\sqrt{-g_E}\Bigg\{
     \frac{M_{\rm P}^2}{2}R_E
                   -\frac{1}{2}g_E^{\mu\nu}(\partial_\mu\psi_E)(\partial_\nu\psi_E)-\frac12g_E^{\mu\nu}(\partial_\mu\phi_E)(\partial_\nu\phi_E)
\label{action dim 4:multiplier:5}\\
     && \hskip 2.2cm
-\frac{M_{\rm P}^2}{2}\Lambda{\rm e}^{4\tilde\psi_E}
 -\frac{m^2}{2}\phi_E^2{\rm e}^{2\tilde\psi_E}-\frac{\lambda}{4!}\phi_E^4
     -\frac{M_{\rm P}^4}{8\alpha}\left[1+\xi\frac{\phi_E^2}{M_{\rm P}^2}
     -{\rm e}^{2\tilde\psi_E}\right]^2
 \Bigg\}
\,,
\nonumber
\end{eqnarray}
where for simplicity we have introduced a rescaled field,
\begin{equation}
  {\tilde \psi}_E =\frac{\psi_E}{\sqrt{6}M_{\rm P}}\equiv \ln(\Omega)
\,.
\label{tilde psiE}
\end{equation}

 We shall now pause to analyze the effective action~(\ref{action dim 4:multiplier:5}) and its link to inflation.
One conventional way of getting inflation is Starobinsky's inflationary model~\cite{Starobinsky:1980te}, 
which is obtained in the limit in which
$\Lambda=m=\lambda=\xi=0$. In that case the effective potential in Eq.~(\ref{action dim 4:multiplier:5})
reduces to, 
\begin{equation}
 (V_E)_{\rm Starobinsky} = \frac{M_{\rm P}^4}{8\alpha}\left[1-{\rm e}^{2\tilde\psi_E}\right]^2
\,.
\label{Starobinsky limit}
\end{equation}
The minimum of the potential, $V_E=0$, is reached when $\tilde\psi_E=\psi_E/[\sqrt{6}M_{\rm P}]=0$. 
The potential~(\ref{Starobinsky limit}) is suitable for (large field) inflation as it exhibits 
a plateau at large negative values of $\psi_E$, {\it i.e.} $\psi_E\ll -M_{\rm P}$, and 
inflationary predictions of the Starobinsky model~\cite{Starobinsky:1980te}
beautifully agree with the measurements~\cite{Ade:2015lrj}, 
provided one chooses $\alpha$ that is consistent 
with the COBE normalization, $\alpha\simeq 1.1\times 10^9$. 
However, as we shall see below, Starobinsky's model is in general unprotected against large corrections arising 
from higher dimensional Planck scale operators.

The other regime is that of Higgs inflation~\cite{Salopek:1988qh,Bezrukov:2007ep}. 
Higgs inflation is obtained in the limit when 
$\lambda\simeq 0.5$, $m\approx 0$,~\footnote{The Higgs mass $m=m_H$ is completely negligible during inflation.
With our convention for $\lambda$, $\lambda=2(m_H/v)^2\simeq 0.5$, where 
$m_H=125~{\rm GeV}$ and $v=246~{\rm GeV}$ are the Higgs mass and its vacuum expectation value, respectively.
The simplistic analysis presented above neglects the running of the Higgs self-coupling $\lambda$,
which is important since it significantly changes with the renormalization group scale. For our purpose this simplified analysis suffices 
as Higgs inflation is not the main focus of our work.
}
 $\alpha=0$, $\Lambda=0$ and $|\xi|\gg 1$, $\xi<0$.
In this regime, the following combination of the two fields,
\begin{equation}
  1+\xi\frac{\phi_E^2}{M_{\rm P}^2} -e^{2\tilde\psi_E} \approx 0 
\,,
\label{massive dof}
\end{equation}
must vanish to a high accuracy  
since the mass associated with that degree of freedom is very large (super-Planckian),
implying that that degree of freedom decouples and becomes non-dynamical.
This means that one is left with only one dynamical degree of freedom.
To get a better grip of that degree of freedom, 
we insert the differential form of Eq.~(\ref{massive dof}) into Eq.~(\ref{action dim 4:multiplier:5}) 
to obtain the following effective Lagrangian for $\psi_E$,
\begin{equation}
 ({\cal L}_E)_{\rm Higgs \; inflation} 
\approx  -\frac12 \left[1-\frac{e^{4\tilde\psi_E}}{6\xi(1-e^{2\tilde\psi_E})}\right]g^{\mu\nu}_E(\partial_\mu \psi_E)(\partial_\nu \psi_E)
-\frac{\lambda M_{\rm P}^4}{4!\xi^2}\left(1-e^{2\tilde\psi_E}\right)^2
\,.
\label{higgs inflation limit}
\end{equation}
This form is still not very useful, since the kinetic term is in a non-canonical form. 
To get it to the canonical form, one would have to make a suitable rescaling of $\psi_E$ which is rather complicated.
Rather than doing that, we note that the plateau of the potential (where the relevant part of inflation happens) 
is attained for large negative values of $\psi_E$,
at which the kinetic term in~(\ref{higgs inflation limit}) becomes approximately canonical 
(this is so because $\exp(4\tilde\psi_E)\rightarrow 0$ in~(\ref{higgs inflation limit})). 
In conclusion, we have thus shown that inflationary predictions of Higgs inflation and Starobinsky's inflation 
are identical, provided one makes the identification, 
\begin{equation}
 \frac{3\xi^2}{\lambda} \leftrightarrow \alpha\simeq 1.1 \times1 0^9
\,.
\label{Higgs versus Starobinsky}
\end{equation}
Of course, the predictions are not exactly identical, because the noncanonical nature of 
the kinetic term in Higgs inflation~(\ref{higgs inflation limit}) will play some role at late stages of inflation, making 
the predictions of two models slightly different. Nevertheless, these differences are tiny and 
the predictions of Higgs inflation lie in the sweet spot 
of the Planck data~\cite{Ade:2015lrj}, just as those of Starobinsky's model. And just as in the case 
of Starobinsky's model, one cannot find a simple argument that would do away the Planck scale operators~\cite{Bezrukov:2010jz},
which will in general spoil the slow roll conditions of Higgs inflation.

Even though we have shown that, when the appropriate limits are taken, the action~(\ref{action dim 4:multiplier:5}) contains 
two very popular and successful inflationary models, 
here we shall analyze the model~(\ref{action dim 4:multiplier:5})
from a different perspective. We want to classify the operators appearing in~(\ref{action dim 4:multiplier:5})
with respect to how they scale with a power of $e^{\tilde \psi_E}$, and we refer to these operators as of 
canonical dimension $d_E$ in the Einstein frame.
There are namely three classes of terms in~(\ref{action dim 4:multiplier:5}),
these that scale as $e^{4\tilde\psi_E}$,  as $e^{2\tilde\psi_E}$ and constant terms. There are no terms that 
scale as a negative power of $e^{2\tilde\psi_E}$ but -- as we shall see in section~\ref{Dimension six operators} 
-- that is simply a consequence of our truncation 
of the action at canonical dimension $d=4$. When operators of higher canonical dimension are included, 
the resulting Einstein frame effective action will contain terms that scale as, $e^{(4-d_E)\tilde\psi_E}$, with 
$d_E/2\in\{0,1,2,\cdots\}\equiv \mathbbm{N}\cup \{0\}$. Because of this property, in the limit of a 
large and positive $\psi_E$ only the operators multiplying $e^{4\tilde\psi_E}$, $e^{2\tilde\psi_E}$ and $e^{0\tilde\psi_E}=1$
will survive; in the limit of a large and negative $\psi_E$,  the operators multiplying $e^{-2n\tilde\psi_E}$, 
with $n=(d_E/2)-2\in\{0,1,2,\cdots\}$ will survive, and there are infinitely many of them. From that observation alone 
we see that our only hope to get rid of an infinite class of operators is to demand that $\psi_E$ 
becomes large and positive (note that both the Higgs and Starobinsky's inflationary model work in the opposite regime).
In that case we need to arrange that the operators multiplying positive powers of $e^{2\tilde\psi_E}$ are 
zero (or finely tuned near zero) that their contribution throughout the history of the Universe is negligible.
Let us now discuss how much fine tuning is required to achieve that.

\medskip

In order to do that analysis consider the Einstein frame action~(\ref{action dim 4:multiplier:5})
that includes operators up to canonical dimension four, $d\leq 4$.
The operators that in the Einstein frame contribute as those of canonical dimension $d_E=0$ to the Lagrangian density are,
\begin{equation}
\Delta_0{\cal L}_E=-\frac{M_{\rm P}^2}{2}\left(\Lambda+\frac{M_{\rm P}^2}{4\alpha}\right){\rm e}^{4\tilde\psi_E}
\,,
\label{canonical dim 0}
\end{equation}
while the operators of canonical dimension $d_E=2$ are,
\begin{equation}
\Delta_2{\cal L}_E=\left[\frac{M_{\rm P}^4}{4\alpha}-\frac12\left(m^2-\frac{\xi M_{\rm P}^2}{2\alpha}\right)\phi_E^2
\right]{\rm e}^{2\tilde\psi_E}
\,.
\label{canonical dim 2}
\end{equation}
From Eq.~(\ref{canonical dim 0}) we see that by choosing the cosmological constant $\Lambda$ such to cancel the latter contribution, {\it i.e.} $\Lambda=-M_{\rm P}^2/(4\alpha)$, removes the problematic dimension zero contribution to the Lagrangian,
resulting in,
\begin{equation}
 \Delta_0{\cal L}_E=0
\,.
\label{L0=0}
\end{equation}
However, the tuning does not completely work for the $d_E= 2$ operator~(\ref{canonical dim 2}). 
Namely, upon choosing
$m^2=\xi M_{\rm P}^2/(2\alpha)$ in Eq.~(\ref{canonical dim 2})
one is still left with the following offending  $d_E= 2$ contribution,
\begin{equation}
\Delta_2{\cal L}_E=\frac{M_{\rm P}^4}{4\alpha}{\rm e}^{2\tilde\psi_E}
\,,
\label{canonical dim 2B}
\end{equation}
which -- as far as we can see -- cannot be removed completely by fine tuning of the parameters in the effective 
action~(\ref{action dim 4:multiplier:5}).~\footnote{One can try to remove that operator by suitably choosing 
$\phi_E$. That is not satisfactory since $\phi_E$ is a dynamical field, and we would like to keep it that way and 
use it if needed for inflation.} Another trouble with that term is that its contribution to the effective potential 
in the Einstein frame is negative for any value of $\psi_E$ and for $\alpha>0$.
One way of arguing away that term is to choose $\alpha$ negative and sufficiently large. As we will see in a moment,
that choice is not acceptable if we are to build inflation from the action~(\ref{action dim 4:multiplier:5}).
Fortunately, this problem is unique to the truncation $d\leq 4$ and 
does not persist when higher dimensional operators are included.

 Assuming that $\Delta_0{\cal L}_E$ and $\Delta_2{\cal L}_E$ can be neglected, 
the remaining terms contributing to the effective potential in~(\ref{action dim 4:multiplier:5})
are those of canonical dimension $d_E=4$,
\begin{equation}
-\Delta_4{\cal L}_E\supset \Delta_4V_E
    =
    \frac{\lambda}{4!}\phi_E^4
     +\frac{M_{\rm P}^4}{8\alpha}\left[1+\xi\frac{\phi_E^2}{M_{\rm P}^2}\right]^2
\,.
\label{canonical dim 4}
\end{equation}
This effective potential can generate a viable inflationary model provided $\lambda$ is small enough and $\alpha$ is large enough.
Indeed, in that case the potential exhibits an approximate plateau, $\Delta_4V_E = M_{\rm P}^4/(8\alpha)$.
The value of $\alpha$ is fixed by the COBE constraint, $\alpha\simeq 1.1\times 10^9$, 
and with $\lambda=0$, $\xi$ negative and 
$|\xi|\ll 1$ (this is the opposite limit from Higgs inflation, see Eq.~(\ref{massive dof})) 
one gets a viable inflationary model
(whose slow roll parameters are to leading order those of the inverted  quadratic model~\cite{Lyth:1998xn}). 
In this model  the inflaton condensate
grows during inflation towards, $1+\xi\phi_E^2/M_{\rm P}^2=0$, at which point
the potential energy minimizes at zero, $\Delta_4V_E=0$ (here we have assumed that 
the contribution from~(\ref{canonical dim 2B}) stays negligibly small during inflation). 
Note that the potential~(\ref{canonical dim 4}) is flat along $\psi_E$, so one could also foresee 
(by a suitable choice of initial conditions)
a period of ultra-slow roll inflation~\cite{Tsamis:2003px,Romano:2015vxz}, followed by
an inverted quadratic inflation along $\phi_E$. 

 As we have seen in this section, including all operators up to the canonical dimension $d=4$ has some undesirable features,
in particular there remains a dimension two operator~(\ref{canonical dim 2B}) that cannot be canceled. For that reason,
to get a more complete understanding of the role of higher dimensional operators during inflation, 
it is necessary to analyze dimension six operators, which is what we do next.

\section{Dimension six operators}
\label{Dimension six operators}

 Even though conceptually straightforward, a complete analysis of dimension six operators is rather tedious, simply 
because there are many of them. For simplicity, in what follows we analyze only the operators that do not 
modify kinetic terms and do not violate unitarity. The $d=6$ operators which contribute to the action are then,
\begin{equation}
 \Delta S^{(6)} = \int d^4x \sqrt{-g} 
\left\{
         \frac{\beta_6}{3} \frac{R^3}{M_{\rm P}^2}
         +\frac{\alpha_6}{2} \frac{R^2\phi^2}{M_{\rm P}^2}
         -\frac{\xi_6}{2} \frac{R\phi^4}{M_{\rm P}^2}
         - \frac{g_6}{6!}\frac{\phi^6}{M_{\rm P}^2}
\right\}
\,.
\label{S6}
\end{equation}
Examples of $d=6$ operators we are not considering are the following kinetic operators, 
$Rg^{\mu\nu}(\partial_\mu\phi)(\partial_\nu\phi)$, 
$\phi^2g^{\mu\nu}(\partial_\mu\phi)(\partial_\nu\phi)$, $G^{\mu\nu}(\partial_\mu\phi)(\partial_\nu\phi)$, {\it etc.},
where $G^{\mu\nu}=R^{\mu\nu} -(1/2)g^{\mu\nu}R$ is the Einstein tensor. One can show that including these operators 
would not in any essential way change our analysis.
In addition, there are $d=6$ operators that violate unitarity, such as $RR_{\mu\nu}R^{\mu\nu}$, $R\Box R$, {\it etc},
which we assume to be absent. As in section~\ref{Dimension four operators}, we can introduce a Lagrange
multiplier field $\omega^2$ that enforces $\Phi=R$. Varying with respect to $\Phi$ gives 
a quadratic equation ({\it cf.} Eq.~(\ref{solution for Phi})),
\begin{equation}
\frac{\Phi^2}{M_{\rm P}^4}+\frac{\alpha+\alpha_6\phi^2/M_{\rm P}^2}{\beta_6}\frac{\Phi}{M_{\rm P}^2}
      +\frac{1}{2\beta_6}\left[
                  1-\left(2\frac{\omega^2}{M_{\rm P}^2}+\xi\frac{\phi^2}{M_{\rm P}^2}+\xi_6\frac{\phi^4}{M_{\rm P}^4}\right)
                                 \right] = 0
\,,
\label{Phi:constraint}
\end{equation}
which is solved by, 
\begin{equation}
\frac{\Phi_\pm}{M_{\rm P}^2}
  = -\frac{\alpha+\alpha_6\phi^2/M_{\rm P}^2}{2\beta_6}
      \pm
   \sqrt{\frac{1}{4\beta_6^2}\left[\alpha+\alpha_6\frac{\phi^2}{M_{\rm P}^2}\right]^2
          -\frac{1}{2\beta_6}\left[1-\left(2\frac{\omega^2}{M_{\rm P}^2}
                               +\xi\frac{\phi^2}{M_{\rm P}^2}+\xi_6\frac{\phi^4}{M_{\rm P}^4}\right) \right]
         }
\,.
  \label{Phi:constraint:solution}
\end{equation}
When transforming to the Einstein frame, just as in section~\ref{Dimension four operators}, one gets,
\begin{equation}
  \omega^2 =\frac{M_{\rm P}^2}{2} e^{-2\tilde\psi_E} 
\,;\quad 
    \tilde\psi_E =\ln(\Omega) \equiv \frac{\psi_E}{\sqrt{6}M_{\rm P}}
\,;\qquad 
     \phi_E = \phi e^{\tilde \psi_E} \equiv M_{\rm P}\tilde \phi_E
\,.
\label{d=6:Einstein frame}
\end{equation}
It is now convenient to rewrite Eq.~(\ref{Phi:constraint:solution}) in the Einstein frame as,
\begin{eqnarray}
\tilde \Phi \equiv \frac{\Phi}{M_{\rm P}^2}
&=&-\frac{\alpha}{2\beta_6}
        -\frac{\alpha_6}{2\beta_6}\tilde \phi_E^2 e^{-2\tilde \psi_E}
\label{Phi:constraint:solution:EF}\\ 
 &+& \frac{{\rm sign}[\alpha]}{2\beta_6}\sqrt{\alpha^2-2\beta_6}
      \sqrt{1+\frac{2}{\alpha^2\!-\!2\beta_6}\left[\beta_6+(\alpha\alpha_6+\beta_6\xi)\tilde\phi_E^2\right]e^{-2\tilde\psi_E}
      + \frac{1}{\alpha^2\!-\!2\beta_6}
             (\alpha_6^2+2\beta_6\xi_6)\tilde\phi_E^4e^{-4\tilde\psi_E}
         }
\,,
\nonumber
\end{eqnarray}
where we have picked the solution that in the limit $\beta_6\rightarrow 0$ reduces to~(\ref{solution for Phi}) 
and we assumed, $\alpha^2-2\beta_6\geq 0$. When expressed in terms of $\tilde\Phi$, the Einstein frame 
action can be written as, 
\begin{eqnarray}
S_E^{(6)}&\equiv & S_E + \Delta S_E^{(6)}= \int d^4x\sqrt{-g_E}\Bigg[
     \frac{M_{\rm P}^2}{2}R_E
                   -\frac{1}{2}g_E^{\mu\nu}(\partial_\mu\psi_E)(\partial_\nu\psi_E)-\frac12g_E^{\mu\nu}(\partial_\mu\phi_E)(\partial_\nu\phi_E)
\label{Einstein frame action:dim 6}\\
    \! &+&\!\frac{M_{\rm P}^4}{2}
\Bigg\{
                            \bigg[
                               \! -\! \frac{\Lambda}{M_{\rm P}^2} + \left(\alpha+\frac{2\beta_6}{3}\tilde\Phi\right)\tilde\Phi^2
                           \bigg]e^{4\tilde\psi_E}
                         +\bigg[
                              \!- \tilde\Phi+ \bigg(\tilde\Phi\big(\!-\xi+\alpha_6\tilde\Phi\big)-\frac{m^2}{M_{\rm P}^2}\bigg)\tilde \phi_E^2
                           \bigg]e^{2\tilde\psi_E}
\nonumber\\
            &&\hskip 1cm
                       -\Big(
                             \frac{\lambda}{12}+\xi_6\tilde\Phi
                           \Big)\tilde \phi_E^4
                          -\frac{g_6}{360}\tilde \phi_E^6e^{-2\tilde\psi_E}
\Bigg\} 
\Bigg]
\nonumber
\end{eqnarray}
In order to get an insight into the scaling of the action~(\ref{Einstein frame action:dim 6}) with $e^{\tilde\psi_E}$, 
it is useful to expand 
$\tilde\Phi$ in~(\ref{Phi:constraint:solution:EF})
in powers of $e^{-2\tilde\psi_E}$ as follows, 
\begin{equation}
  \tilde\Phi = \tilde\Phi_0 + \tilde\Phi_{-2}e^{-2\tilde\psi_E} + \tilde\Phi_{-4}e^{-4\tilde\psi_E} + \tilde\Phi_{-6}e^{-6\tilde\psi_E}
               +{\cal O}\left(e^{-8\tilde\psi_E}\right)
\,,
\label{expansion of tilde Phi}
\end{equation}
where 
\begin{eqnarray}
\tilde \Phi_0 &=& -\frac{{\rm sign}[\alpha]}{2\beta_6}\left[|\alpha|-\sqrt{\alpha^2\!-\!2\beta_6}\right] 
\label{Phi0}\\
\tilde \Phi_{-2} &=& -\frac{\alpha_6}{2\beta_6}\tilde\phi_E^2
                   +\frac{{\rm sign}[\alpha]}{ 2\beta_6\sqrt{\alpha^2\!-\!2\beta_6} }\left\{\beta_6
                              +(\alpha\alpha_6+\beta_6\xi)\tilde\phi_E^2\right\}
\label{Phi-2}\\
\tilde \Phi_{-4} &=& 
                   \frac{{\rm sign}[\alpha]}{4\beta_6(\alpha^2\!-\!2\beta_6)^{3/2} }\left\{
                              -\big[\beta_6+(\alpha\alpha_6+\beta_6\xi)\tilde\phi_E^2\big]^2
                             +(\alpha^2\!-\!2\beta_6) \big[\alpha_6^2+2\beta_6\xi_6\big]\tilde\phi_E^4\right\}
\label{Phi-4}\\
\tilde \Phi_{-6} &=& 
                   \frac{{\rm sign}[\alpha]}{4\beta_6(\alpha^2\!-\!2\beta_6)^{5/2} }\big[\beta_6+(\alpha\alpha_6+\beta_6\xi)\tilde\phi_E^2\big]
                        \left\{
                              \big[\beta_6+(\alpha\alpha_6+\beta_6\xi)\tilde\phi_E^2\big]^2
                             -(\alpha^2\!-\!2\beta_6) \big[\alpha_6^2+2\beta_6\xi_6\big]\tilde\phi_E^4\right\}
\,.
\label{Phi-6}
\end{eqnarray}
Upon inserting~(\ref{expansion of tilde Phi}) into the action~(\ref{Einstein frame action:dim 6}), one can 
resolve terms multiplying different powers of $e^{2\tilde\psi_E}$. 
The fastest growing terms are the terms multiplying $e^{4\tilde\psi_E}$. They contribute 
to the Lagrangian density as the cosmological constant, and hence we demand that to a high accuracy those terms vanish,
\begin{equation}
\Delta_4 {\cal L} = \frac{M_{\rm P}^4}{2}
       \bigg[
                               \! -\! \frac{\Lambda}{M_{\rm P}^2} 
                                + \left(\alpha+\frac{2\beta_6}{3}\tilde\Phi_0\right)\tilde\Phi_0^2
                           \bigg]e^{4\tilde\psi_E} \approx 0 
\,.
\label{terms exp 4 tilde psiE}
\end{equation}
This can be achieved {\it e.g.} by the appropriate choice of $\Lambda$. This fine tuning corresponds 
to the usual tuning of the cosmological constant, and we have nothing new to add here. 
The second set of terms are those multiplying $e^{2\tilde\psi_E}$,
\begin{equation}
\Delta_2 {\cal L} =\frac{M_{\rm P}^4}{2}\bigg\{\! - \tilde\Phi_{0}
                              +2\left(\alpha+\beta_6\tilde\Phi_0\right)\tilde \Phi_0 \tilde\Phi_{-2}
    +\bigg[
                               \tilde\Phi_0\left(-\xi+\alpha_6\tilde\Phi_0\right)-\frac{m^2}{M_{\rm P}^2}
                           \bigg]\tilde \phi_E^2\bigg\}e^{2\tilde\psi_E}
 \approx 0 
\,.
\label{terms exp 4 tilde psiE}
\end{equation}
As indicated in this equation, we also demand that this term vanishes to a sufficiently high accuracy.
As Eq.~(\ref{terms exp 4 tilde psiE}) contains two types of terms -- 
those that do not contain any power of $\tilde \phi_E$ and those multiplying $\tilde \phi_E^2$
-- that amounts to two more fine tunings of the parameters.
We have quite a  few parameters at the disposal 
(in that sense the situation becomes better when more higher dimensional operators are 
included). For example, a suitable choice of the mass term $m^2$ 
and of $\beta_6$ can do the job
(the COBE normalization is now set by $\alpha^2-2\beta_6$, so we can choose at will either 
$\alpha$ or $\beta_6$ to fine tune the $d_E=4$ term). 
Assuming that the terms that contain a positive power of $e^{2\tilde\psi_E}$
are so fine tuned that they can be neglected, 
we are left with the constant term ($d_E=4$) and with the terms that scale with negative powers of $e^{2\tilde\psi_E}$
($d_E\geq 6$),
{\it i.e.} the terms that decay for large values of $\tilde\psi_E$. 
Demanding a sufficiently large $\psi_E$ can render all higher order terms irrelevant, such that during inflation 
only the constant term (and possibly the term $\propto e^{-2\tilde\psi_E}$) are of any importance. 
 The terms that remain in the Einstein action~(\ref{Einstein frame action:dim 6}) are to a good approximation,
\begin{eqnarray}
S_E^{(6)}&\approx& \int d^4x\sqrt{-g_E}\Bigg\{
     \frac{M_{\rm P}^2}{2}R_E
                   -\frac{1}{2}g_E^{\mu\nu}(\partial_\mu\psi_E)(\partial_\nu\psi_E)-\frac12g_E^{\mu\nu}(\partial_\mu\phi_E)(\partial_\nu\phi_E)
\label{Einstein frame action:dim 6:B}\\
    \! &+&\!\frac{M_{\rm P}^4}{2}
                            \bigg[
                            \! -\tilde\Phi_{-2}+\alpha\Big(\tilde \Phi_{-2}^2+2\tilde \Phi_{0}\tilde \Phi_{-4}\Big)
                                           +2\beta_6\tilde \Phi_{0}\Big(\tilde \Phi_{0}\tilde \Phi_{-4}+\tilde \Phi_{-2}^2\Big)
                         +\tilde\Phi_{-2}\left(-\xi+2\alpha_6\tilde\Phi_{0}\right)\tilde \phi_E^2
                       -\Big(
                             \frac{\lambda}{12}+\xi_6\tilde\Phi_0
                           \Big)\tilde \phi_E^4
\bigg]
\nonumber\\
   \! &+&\!\frac{M_{\rm P}^4}{2}
                            \bigg[
                             \tilde\Phi_{-4}\!+\!2\alpha\Big(\tilde \Phi_{0}\tilde \Phi_{-6}\!+\!\tilde \Phi_{-2}\tilde \Phi_{-4}\Big)
              +2\beta_6\Big(\tilde \Phi_{0}^2\tilde \Phi_{-6}\!+\!2\tilde \Phi_{0}\tilde \Phi_{-2}\tilde \Phi_{-4}
                               \!+\!\frac13\tilde \Phi_{-2}^3\Big)
\nonumber\\
  &&\hskip 0.8cm        +\left(-\xi\tilde\Phi_{-4}+\alpha_6\big(2\tilde\Phi_0\tilde\Phi_{-4}\!+\!\tilde\Phi_{-2}^2\big)\right)\tilde \phi_E^2
                       -\xi_6\tilde\Phi_{-2}\tilde \phi_E^4-\frac{g_6}{360}\tilde \phi_E^6
\bigg]e^{-2\tilde\psi_E}
\Bigg\} 
,
\nonumber
\end{eqnarray}
where the second line contains the $d_E=4$ terms that are independent on $\psi_E$ and  the last two lines 
contain the $d_E=6$ terms that scale as $e^{-2\tilde\psi_E}$.
  One can show that a similar analysis goes through when other (kinetic) operators of dimension six are included.
More importantly, an analogous analysis can be carried through when even higher dimensional operators are included 
(albeit the analysis becomes more technically involved). 
One can easily convince oneself that, truncating at an arbitrary but finite order $d\geq 6$, 
in general three fine tunings suffice to get rid of all terms that contain positive powers of $e^{2\tilde\psi_E}$. In that 
sense the analysis presented in this section is generic. It would be incorrect to think that, every time 
one adds new higher dimensional operators, one ought to re-tune. Nature has chosen (unknown) operators that 
carry information about the Planck scale physics, and therefore fine tuning needs to be done only once for the set
 of operators that Nature has picked.

The next question we need to address is whether the action~(\ref{Einstein frame action:dim 6:B}) is suitable
for inflationary model building.
The analysis of inflation 
 is in fact very similar to that  for dimension four 
effective potential~(\ref{canonical dim 4}) we do in section~\ref{Dimension four operators}
and therefore here we only present an analysis in broad brush strokes.
From the current CMB observations we know that the constant term the second line in~(\ref{Einstein frame action:dim 6:B})
ought to be tuned to accord with the COBE normalization,
and slow rolling can be either along the $\phi_E$ direction (which corresponds to a decrease in potential energy) 
and/or along an increasing $\psi_E$. 
Depending on what the values of the masses of the two scalar fields are, 
inflation will either occur in the $\psi_E$ direction (when the scalaron is the lighter field) 
or in the scalar field direction (when $\phi_E$ is the lighter field). When the masses of  the two fields are comparable, 
one can gets a genuine two-field inflationary model.
In the former case (rolling along $\psi_E$) one gets a Starobinsky-like inflationary model, with the important difference 
that the scalaron rolls in the direction of an increasing scalaron condensate. That is not a problem since the evolution of 
the scalaron and Ricci scalar become decoupled in this model since the scalaron and graviton become
independent degrees of freedom in the limit when $\psi_E\rightarrow \infty$.
In the latter case (when $\phi_E$ is rolling) inflation can be similar to that of an inverted quadratic potential.
Finally we mention that one can start inflation with  non-attractor initial conditions that at early stages yield
an ultra-slow roll inflation along $\psi_E$, which at a later stage turns into a slow-roll inflation along $\phi_E$.

\section{Conclusions and discussion}
\label{Conclusions and discussion}

 In this paper we have studied the role of Planck scale operators the arise as threshold corrections from 
the unknown Planck scale physics. The central result of this work is:
\begin{itemize}
\item[]
{\it Provided one permits a reasonable amount of fine tuning (precisely three fine tunings are needed), 
one can get a flat enough effective potential in the Einstein frame
to grant inflation whose predictions are consistent with observations.} 
\end{itemize}
This remains true independently on how many Planck scale operators one adds (provided there are sufficiently many of them).
Canonical quantization of the effective theory in general results in changed values of the couplings for each of the operators, 
but does not in any essential way change the conclusions reached in this work. 

 Our hope (and the principal motivation for this work) is that our insights will help to advance the understanding 
of quantum (loop) corrections during inflation. In particular, we hope we will be able to understand whether, how 
and under what conditions these quantum effects can become detectable.

\section*{Acknowledgemetns}

 This work is in part supported by the D-ITP consortium, a program of the Netherlands Organization for
Scientific Research (NWO) that is funded by the Dutch Ministry of Education, Culture and Science
(OCW). A.M. is funded by NEWFELPRO, an International Fellowship Mobility Programme for
Experienced Researchers in Croatia and by the D-ITP.

\end{document}